\documentclass{article}

\usepackage{arxiv}

\usepackage[utf8]{inputenc} % allow utf-8 input
\usepackage[T1]{fontenc}    % use 8-bit T1 fonts
\usepackage[
bookmarks=true,
hypertexnames=true,
hyperindex=true,
hyperfootnotes=true,
linktocpage=false,
hyperfigures=true,
breaklinks=true, 
pdfborder={0 0 0},
colorlinks=true,
linkcolor=black,
unicode=true,						
citecolor=blue,
urlcolor=blue,
pdfinfo={
Title={A GeoSPARQL Compliance Benchmark},
Author={Milos Jovanovik, Timo Homburg, Mirko Spasić},
Keywords={GeoSPARQL, Benchmarking, RDF, SPARQL}
}
]{hyperref}
\usepackage{url}            % simple URL typesetting
\usepackage{booktabs}       % professional-quality tables
\usepackage{amsfonts}       % blackboard math symbols
\usepackage{nicefrac}       % compact symbols for 1/2, etc.
\usepackage{microtype}      % microtypography
\usepackage{lineno}
\usepackage{graphicx}
\usepackage{verbatim}
\usepackage{todonotes}
\usepackage{listingsutf8}
\usepackage{tikz}
\usepackage{xcolor}
\usepackage{rotating}
\usepackage{rotfloat}
\usepackage{cleveref}

\newcommand{\reqinit}{
    \newcounter{reqcountbackup}
    \newcounter{reqcount}
    \renewcommand{\thereqcount}{\textbf{Req.~\arabic{reqcount}}:}
}

\newcommand{\reqstart}{
    \begin{list}{\thereqcount}{\usecounter{reqcount}}
    \setcounter{reqcount}{\value{reqcountbackup}}
}

\newcommand{\reqend}{
    \setcounter{reqcountbackup}{\value{reqcount}}
    \end{list}
}

\title{A GeoSPARQL Compliance Benchmark}

\author{
  Milos Jovanovik\\
  Ss. Cyril and Methodius Univesity in Skopje, N. Macedonia\\
  OpenLink Software, London, UK\\
  \texttt{milos.jovanovik@finki.ukim.mk} \\
  \And
  Timo Homburg\\
  Mainz University Of Applied Sciences, Germany\\
  \texttt{timo.homburg@hs-mainz.de} \\
  \And
  Mirko Spasi\'c\\
  University of Belgrade, Serbia\\
  OpenLink Software, London, UK\\
  \texttt{mirko@matf.bg.ac.rs}
}
\reqinit
\begin{document}
\maketitle

\begin{abstract}
We propose a series of tests that check for the compliance of RDF triplestores with the GeoSPARQL standard. The purpose of the benchmark is to test how many of the requirements outlined in the standard a tested system supports and to push triplestores forward in achieving a full GeoSPARQL compliance. This topic is of concern because the support of GeoSPARQL varies greatly between different triplestore implementations, and such support is of great importance for the domain of geospatial RDF data. Additionally, we present a comprehensive comparison of triplestores, providing an insight into their current GeoSPARQL support.
\end{abstract}


\section{Introduction}
\label{sec:introduction}

The geospatial Semantic Web \cite{fonseca2002geospatial} as part of the Semantic Web \cite{berners2001semantic} represents an ever-growing semantically interpreted wealth of geospatial information. The initial research \cite{battle2011geosparql} and the subsequent introduction of the OGC GeoSPARQL standard \cite{open2012ogc} formalized geospatial vector data representations (WKT \cite{herring2011opengis} and GML \cite{portele2012ogc}) in ontologies, and extended the SPARQL query language \cite{prudhommeaux2008sparql} with support for spatial relation operators. 

Several RDF storage solutions have since adopted GeoSPARQL to various extents as features of their triplestore implementations \cite{battle2012enabling,albiston2018geosparql}. These varying levels of implementation may lead to some false assumptions of users when choosing an appropriate triplestore implementation for their project. For example, some implementations allow for defining a coordinate reference system (CRS) \cite{tobin1964coordinate} in a given WKT geometry literal as stated in the GeoSPARQL standard (e.g.~GraphDB). Other implementations do not allow a CRS definition and instead only support the world geodetic system WGS84 (e.g.~RDF4J) \cite{decker1986world}. Such implementations, even though incomplete according to the GeoSPARQL standard, still cover many geospatial use-cases and can be useful in many scenarios. But, they are not useful, for example, for a geospatial authority that needs to work with many different coordinate system definitions.

The requirements of GeoSPARQL compliant triplestores have been clearly spelled out in the GeoSPARQL standard \cite{open2012ogc}. However, the Semantic Web and GIS community lack a compliance test suite for GeoSPARQL, which we contribute in this publication. We hope that our contribution may be added to the list of OGC conformance tests\footnote{OGC Test Suites: \href{https://cite.opengeospatial.org/teamengine/}{https://cite.opengeospatial.org/teamengine/}}, as they lack a suitable test suite for GeoSPARQL.

Our paper is organized as follows. In \Cref{sec:relatedwork} we discuss existing approaches that worked towards evaluating geospatial triplestores, \Cref{sec:benchmark} introduces the test framework of the benchmark and describes how the compliance tests were implemented. \Cref{sec:experimentalsetup} describes the application of the defined test framework against different triplestore implementations, and we discuss the results in \Cref{sec:results}. In \Cref{sec:limitations} we lay out the limitations of our approach, before concluding the work in \Cref{sec:conclusions}.

\section{Related Work}
\label{sec:relatedwork}

Most standards define requirements which need to be fulfilled to satisfy the standard definition. However, not all standards expose explicit descriptions on how to test compliance with their requirements or a test suite that tests the overall compliance to the standard.

%\todo[inline]{These first two subsections seems a bit misplaced. Maybe we could merge them with the section of Performance vs. Compliance Benchmarks?}

%\subsection{GeoSPARQL}
%\label{sec:geosparql}

GeoSPARQL \cite{open2012ogc}, as an extension of the SPARQL \cite{prudhommeaux2008sparql} query language, defines an ontology model to represent vector geometries, their relations and serializations in WKT and GML, a set of geometry filter functions, an RDFS entailment extension, a query rewrite extension to simplify geospatial queries and further geometry manipulation functions. 

%\subsection{Performance Benchmarks vs. Compliance Benchmarks}
%\label{sec:performancebenchmarks}

First, it is important that we distinguish between performance benchmarks and compliance benchmarks. Performance benchmarks try to evaluate the performance of system, usually by employing a set of queries. Performance benchmarks may also consider semantically equivalent implementations that are not following the syntax specified by a given standard. On the other hand, compliance benchmarks are not concerned with the efficiency or overall performance of a system, but rather with its ability to fulfill certain requirements.

%\subsection{Compliance Benchmark Implementations}
%\label{sec:sparqlbenchmarks}

Several benchmark implementations targeting geospatial triplestores, such as the Geographica Series \cite{garbis2013geographica,ioannidis2019evaluating} or \cite{albiston2018geosparql}, try to evaluate the performance of geospatial function implementations. Both approaches originate from the Linked Data community. Additionally, \cite{huang2019assessment} shows that the geospatial community is interested in benchmarking geospatial triplestores, as well. Their benchmark includes a newly created dataset and tests GeoSPARQL filter functions. While the aforementioned benchmarks might reveal if functions are implemented, they do not necessarily reveal an incorrect implementation of a given function. 

The Tests for Triplestores (TFT) benchmark \cite{rafes2014tft} includes a GeoSPARQL subtest. However, the subtest used here is based on the six example SPARQL queries and the example dataset defined in Annex B of the GeoSPARQL standard \cite{open2012ogc}. Although these examples are a good starting point, they are of informative nature and are intended as guidelines. Therefore, any benchmark based solely on them does not even begin to cover all possible requirements or the multiple ways in which they have to be tested, in order for a system to be deemed as compliant with the standard.

Recently, the EuroSDR group reused the benchmark implementation of \cite{huang2019assessment} to implement a small GeoSPARQL compliance benchmark\footnote{EuroSDR GeoSPARQL Test: \href{https://data.pldn.nl/eurosdr/geosparql-test}{https://data.pldn.nl/eurosdr/geosparql-test}}. This compliance benchmark consists of 27 queries testing a selection of GeoSPARQL functions on a test dataset. In contrast to our benchmark, this implementation does not explicitly test all requirements defined in the GeoSPARQL standard. In particular, GML support, RDFS entailment support and the query rewrite extension, among others, have not been tested in this benchmark.

\section{GeoSPARQL Compliance Benchmark}
\label{sec:benchmark}

The GeoSPARQL compliance benchmark is based on the requirements defined in the GeoSPARQL standard \cite{open2012ogc}. The 30 requirements defined in the standard are grouped into 6 categories and refer to the core GeoSPARQL ontology model and a set of extensions which systems need to implement, and which need to be tested in our benchmark:

\begin{enumerate}
    \item Core component (CORE): Defines the top-level spatial vocabulary components (Requirements 1 - 3)
    \item Topology vocabulary extension (TOP): Defines the topological relation vocabular (Requirements 4 - 6)
    \item Geometry extension (GEOEXT): Defines the geometry vocabulary and non-topological query functions (Requirements 7 - 20)
    \item Geometry topology extension (GTOP): Defines topological query functions for geometry objects (Requirements 21 - 24)
    \item RDFS entailment extension (RDFSE): Defines a mechanism for matching implicit (inferred) RDF triples that are derived based on RDF and RDFS semantics, i.e. derived from RDFS reasoning (Requirements 25 - 27)
    \item Query rewrite extension (QRW): Defines query transformation rules for computing spatial relations between spatial objects based on their associated geometries (Requirements 28 - 30)
\end{enumerate}

Each of the specified requirements may be tested using a set of guidelines which are loosely defined in the abstract test suite in Annex A of the GeoSPARQL standard \cite{open2012ogc}. While the abstract test suite defines the test purpose, method and type to verify if a specific requirement has been fulfilled, it does not define a concrete set of SPARQL queries and a test dataset which may be used for reference. We contribute the test dataset and the set of SPARQL queries to verify each requirement in this publication.

In the GeoSPARQL compliance benchmark, each requirement is tested by one or more SPARQL queries, where there is a single expected answer or a set of expected answers. The number of queries used to test a requirement, as well as the number of expected answers per query, depends on the nature of the requirement. For some of them, it is sufficient to have a single query and a single expected answer to test whether the system under test complies with it. In contrast, other requirements have sub-requirements -- for example, requirements which refer to multiple properties or functions, requirements referring to functions which can be used with geometries with different serializations, or requirements which need a broader coverage of cases, to make sure they are fully met. In these cases, multiple queries are used. Multiple logically equivalent expected answers are used when the answer of a SPARQL query can be technically expressed in different formats or literal serializations.

This approach of using queries and expected answers as tests, allows us to measure the compliance of any RDF storage system by using the HOBBIT benchmarking platform \cite{ngomo2016hobbit, roder2019hobbit}.

The output of the benchmark is a percentage which measures the overall compliance of the tested system with the GeoSPARQL standard. It measures the number of supported requirements of the system, out of the 30 specified requirements, as a percentage.

\subsection{Benchmark Dataset}
\label{sec:benchmarkdataset}

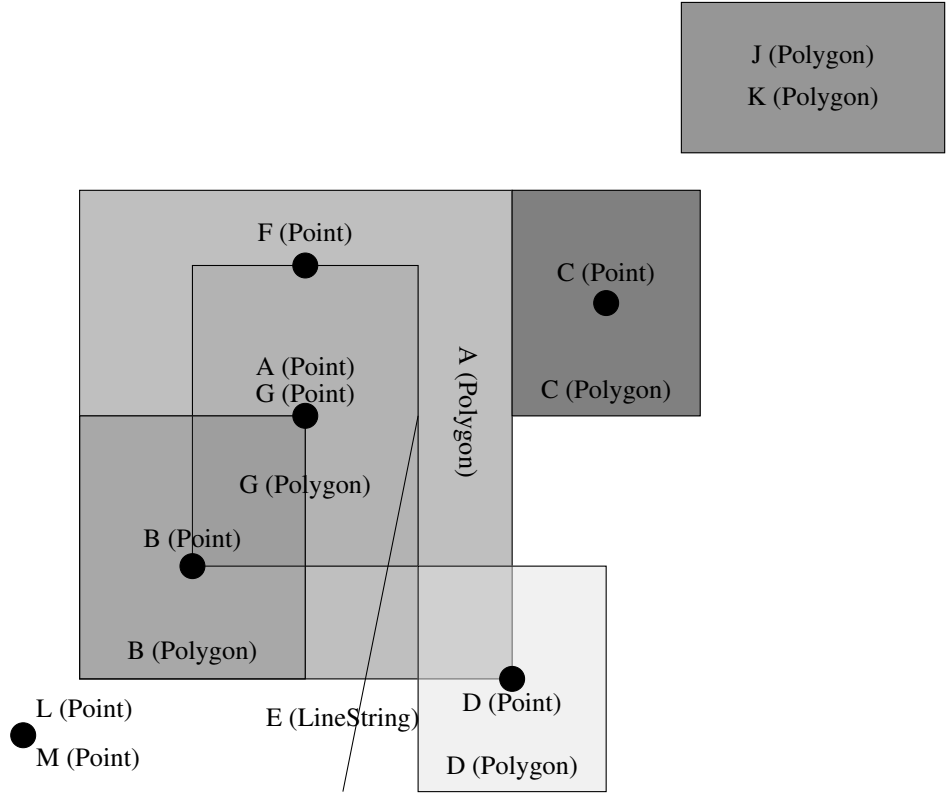
\begin{figure}[!ht]
    \centering
    \begin{tikzpicture}
    \selectcolormodel{gray}
    \fill [draw=black,color=lightgray] (-0.5,-0.5) rectangle (5.25,6.0);
    \draw [draw=black] (-0.5,-0.5) rectangle (5.25,6.0);
    \node[above right=2pt of {(4.3,4.0)},rotate=-90] {A (Polygon)};
    \fill [draw=black,color=orange,fill opacity=0.3] (1.0,1.0) rectangle (4.0,5.0);
    \draw [draw=black] (1.0,1.0) rectangle (4.0,5.0);
    \fill [draw=black,color=red,fill opacity=0.2] (-0.5,-0.5) rectangle (2.5,3.0);
    \draw [draw=black] (-0.5,-0.5) rectangle (2.5,3.0);
    \node[above=2pt of {(2.5,1.7)}] {G (Polygon)};
    \node[above=2pt of {(1.0 ,-0.5)}] {B (Polygon)};
    \node [draw, circle,fill ] at (1.0 ,1.0) {}; 
    \node[above=2pt of {(1.0 ,1.0)}] {B (Point)};
    \fill [draw=black,color=gray] (7.75,3.0) rectangle (5.25,6.0);
    \draw [draw=black] (7.75,3.0) rectangle (5.25,6.0);
    \node[above=1pt of {(6.5,3.0)}] {C (Polygon)};
    \node [draw , circle, fill ] at (6.5 ,4.5) {};
    \node[above=3pt of {(6.5,4.5)}] {C (Point)};
    \fill [draw=black,color=yellow,fill opacity=0.5] (6.5,1.0) rectangle (4.0,-2.0);
    \draw [draw=black] (6.5,1.0) rectangle (4.0,-2.0);
    \node[above=1pt of {(5.25,-2.0)}] {D (Polygon)};
    \node [draw , circle,fill] at (5.25 ,-0.5) {}; 
    \node[below=2pt of {(5.25 ,-0.5)}] {D (Point)};
    \draw (3.0,-2.0) -- (4.0, 3.0);
    \node[above=5pt of {(3.0,-1.5)}] {E (LineString)};
    \node [draw , circle, fill ] at (2.5 ,5.0) {};
    \node[above=4pt of {(2.5 ,5.0)}] {F (Point)};
    \node [draw , circle, fill] at (2.5 ,3.0) {}; 
    \node[above=10pt of {(2.5 ,3.0)}] {A (Point)};
    \node[above=0pt of {(2.5 ,3.0)}] {G (Point)};
    \fill [draw=black,color=green] (7.5,6.5) rectangle (11.0,8.5) ;
    \draw [draw=black] (7.5,6.5) rectangle (11.0,8.5) ;
    \node[above=0pt of {(9.25,7.5)}] {J (Polygon)};
    \node[below=0pt of {(9.25,7.5)}] {K (Polygon)};
    \node[above right=2pt of {(-1.25 ,-1.25)}] {L (Point)};
    \node[below right=2pt of {(-1.25 ,-1.25)}] {M (Point)};
    \node [draw , circle,fill ] at (-1.25 ,-1.25) {}; 
    \end{tikzpicture}
    \caption{Abstract view of the geometries which are part of the benchmark dataset. Geometries A, B, C, D, G, J and K represent \texttt{Polygon} geometries and (aside from J and K) all have a center \texttt{Point} geometry, as well. Geometry E represents a \texttt{LineString} geometry, while geometries F, L and M represent \texttt{Point} geometries. Geometries H and I are empty geometries and not visible in this figure. All geometries are represented in the CRS84 geodetic system, except for geometry M which is represented in EPSG:4326. Each geometry is represented both using WKT and GML literals.}
    \label{fig:geometrydataset}
\end{figure}

The GeoSPARQL standard defines an example dataset for testing in its Annex B \cite{open2012ogc}, which can be used with the set of 6 example test queries defined in the same section. This example dataset contains 6 geometries. We wanted to use this dataset, but given that we aimed to test all requirements of the standard, we had to substantially extend the dataset both with new geometries and additional properties of the existing geometries. \Cref{fig:geometrydataset} shows the geometries included in our extended dataset.

The extended benchmark dataset contains 13 geometries of \texttt{Polygon}, \texttt{Point} and \texttt{LineString} types, all expressed as both WKT and GML literals. The total size of the RDF dataset is over 300 triples. The dataset is available as part of the benchmark code \cite{geosparqlcompliancebenchmarkgithub}, in RDF/XML, GeoJSON \cite{butler2014geojson} and GML representations.

\subsection{Benchmark Queries}
\label{benchmarkqueries}

We provide here an overview of the approach we had in writing the queries used by the benchmark to test the requirements of the GeoSPARQL standard. The requirements are presented in order of the GeoSPARQL extension definitions presented in \Cref{sec:benchmark}. The benchmark queries are available as part of the benchmark code \cite{geosparqlcompliancebenchmarkgithub}. The details about how each test and sub-test is scored are presented in \Cref{sec:benchmarkresults}.

\reqstart
%\begin{itemize}
    \item Implementations shall support the SPARQL Query Language for RDF \cite{prudhommeaux2008sparql}, the SPARQL Protocol for RDF \cite{clark2008sparql} and the SPARQL Query Results XML Format \cite{beckett2008sparql}.
%\end{itemize}
\reqend

We test requirement 1 with a single, basic SPARQL query which selects the first triple where geometry A is the subject. To get consistent results across different systems, we have to use a specific subject and have to order the results.

%\begin{itemize}
\reqstart
    \item Implementations shall allow the RDFS \cite{brickley1999resource} class \href{http://www.opengis.net/ont/geosparql#SpatialObject}{\texttt{geo:SpatialObject}} to be used in SPARQL graph patterns.
    \item Implementations shall allow the RDFS class \href{http://www.opengis.net/ont/geosparql#Feature}{\texttt{geo:Feature}} to be used in SPARQL graph patterns.
    \label{req:1}
\reqend
%\end{itemize}

Requirements 2 and 3 are tested with single SPARQL queries, which select the first entity of type \href{http://www.opengis.net/ont/geosparql#SpatialObject}{\texttt{geo:SpatialObject}} and \href{http://www.opengis.net/ont/geosparql#Feature}{\texttt{geo:Feature}}, respectively. In order to get consistent results for both queries across different systems, we order the results.

\reqstart
%\begin{itemize}
    \item Implementations shall allow the properties \href{http://www.opengis.net/ont/geosparql#sfEquals}{\texttt{geo:sfEquals}}, \href{http://www.opengis.net/ont/geosparql#sfDisjoint}{\texttt{geo:sf\-Dis\-joint}}, \href{http://www.opengis.net/ont/geosparql#sfIntersects}{\texttt{geo:sfIntersects}}, \href{http://www.opengis.net/ont/geosparql#sfTouches}{\texttt{geo:sfTouches}}, \href{http://www.opengis.net/ont/geosparql#sfCrosses}{\texttt{geo:sfCrosses}}, \href{http://www.opengis.net/ont/geosparql#sfWithin}{\texttt{geo:sf\-With\-in}}, \href{http://www.opengis.net/ont/geosparql#sfContains}{\texttt{geo:sfContains}}, \href{http://www.opengis.net/ont/geosparql#sfOverlaps}{\texttt{geo:sfOverlaps}} to be used in SPARQL graph patterns.
    \item Implementations shall allow the properties \href{http://www.opengis.net/ont/geosparql#ehEquals}{\texttt{geo:ehEquals}}, \href{http://www.opengis.net/ont/geosparql#ehDisjoint}{\texttt{geo:eh\-Dis\-joint}}, \href{http://www.opengis.net/ont/geosparql#ehMeet}{\texttt{geo:ehMeet}}, \href{http://www.opengis.net/ont/geosparql#ehOverlap}{\texttt{geo:ehOverlap}}, \href{http://www.opengis.net/ont/geosparql#ehCovers}{\texttt{geo:ehCovers}}, \href{http://www.opengis.net/ont/geosparql#ehCoveredBy}{\texttt{geo:eh\-Covered\-By}}, \href{http://www.opengis.net/ont/geosparql#ehInside}{\texttt{geo:ehInside}}, \href{http://www.opengis.net/ont/geosparql#ehContains}{\texttt{geo:ehContains}} to be used in SPARQL graph patterns.
    \item Implementations shall allow the properties \href{http://www.opengis.net/ont/geosparql#rcc8eq}{\texttt{geo:rcc8eq}}, \href{http://www.opengis.net/ont/geosparql#rcc8dc}{\texttt{geo:rcc8dc}}, \href{http://www.opengis.net/ont/geosparql#rcc8ec}{\texttt{geo:rcc8ec}}, \href{http://www.opengis.net/ont/geosparql#rcc8po}{\texttt{geo:rcc8po}}, \href{http://www.opengis.net/ont/geosparql#rcc8tppi}{\texttt{geo:rcc8tppi}}, \href{http://www.opengis.net/ont/geosparql#rcc8tpp}{\texttt{geo:rcc8tpp}}, \href{http://www.opengis.net/ont/geosparql#rcc8ntpp}{\texttt{geo:rcc8ntpp}}, \href{http://www.opengis.net/ont/geosparql#rcc8ntppi}{\texttt{geo:rcc8ntppi}} to be used in SPARQL graph patterns.
%\end{itemize}
\reqend

We test each of the requirements 4, 5 and 6 with eight different queries, to test the sub-requirements for each property specified. Since the queries for requirements 28, 29 and 30 require the use of these same properties to test the system's compliance to the GeoSPARQL RIF \cite{kifer2013rif} rules, we use an approach where the explicit RDF triples needed to test requirements 4, 5 and 6 involve geometries which are the top result when using the ordering of the query results. For this purpose, the queries for requirements 4, 5, and 6 order the results, and select the top result only, to ensure they test the existence of the explicit and materialized RDF triple in the dataset.

\reqstart
%\begin{itemize}
    \item Implementations shall allow the RDFS class \href{http://www.opengis.net/ont/geosparql#Geometry}{\texttt{geo:Geometry}} to be used in SPARQL graph patterns.
    \item Implementations shall allow the properties \href{http://www.opengis.net/ont/geosparql#hasGeometry}{\texttt{geo:hasGeometry}} and \href{http://www.opengis.net/ont/geosparql#hasDefaultGeometry}{\texttt{geo\-:\-has\-Default\-Geometry}} to be used in SPARQL graph patterns.
    \item Implementations shall allow the properties \href{http://www.opengis.net/ont/geosparql#dimension}{\texttt{geo:dimension}}, \href{http://www.opengis.net/ont/geosparql#coordinateDimension}{\texttt{geo:\-co\-ordinate\-Dimension}}, \href{http://www.opengis.net/ont/geosparql#spatialDimension}{\texttt{geo:spatialDimension}}, \href{http://www.opengis.net/ont/geosparql#isEmpty}{\texttt{geo:isEmpty}}, \href{http://www.opengis.net/ont/geosparql#isSimple}{\texttt{geo:\-is\-Sim\-ple}}, \href{http://www.opengis.net/ont/geosparql#hasSerialization}{\texttt{geo:hasSerialization}} to be used in SPARQL graph patterns.
%\end{itemize}
\reqend

The tests for requirements 7, 8 and 9 are done by selecting all entities of type \href{http://www.opengis.net/ont/geosparql#Geometry}{\texttt{geo:Geometry}} (Req. 7), or by selecting the object/value of geometry A denoted by the property in question (Req. 8 and 9). Since requirement 8 specifies two distinct properties, and requirement 9 specifies six such properties, the tests for these requirements consist of two and six queries, respectively.

\reqstart
%\begin{itemize}
    \item All RDFS Literals of type \href{http://www.opengis.net/ont/geosparql#wktLiteral}{\texttt{geo:wktLiteral}} shall consist of an optional URI identifying the coordinate reference system followed by Simple Features Well Known Text (WKT) describing a geometric value. Valid \href{http://www.opengis.net/ont/geosparql#wktLiteral}{\texttt{geo:wktLiteral}} instances are formed by concatenating a valid, absolute URI as defined in \cite{berners1998rfc2396}, one or more spaces (Unicode U+0020 character) as a separator, and a WKT string as defined in Simple Features \cite{iso119125}.
%\end{itemize}
\reqend

We test requirement 10 by selecting and checking the datatype of a correctly defined WKT literal from the dataset, to make sure the system under test supports the specified format of WKT literals and their datatype.

\reqstart
%\begin{itemize}
    \item URI \href{http://www.opengis.net/def/crs/OGC/1.3/CRS84}{\texttt{<http://www.opengis.net/def/crs/OGC/1.3/CRS84>}}
    shall be assumed as the spatial reference system for \href{http://www.opengis.net/ont/geosparql#wktLiteral}{\texttt{geo:wktLiterals}} that do not specify an explicit spatial reference system URI.
%\end{itemize}
\reqend

We test requirement 11 by first defining two geometries in the dataset: J and K, which represent the same polygon, but geometry K has a WKT literal with an explicitly specified reference system, while geometry J does not contain the URI and only contains the polygon points in the literal value: 

\begin{verbatim}
J: Polygon((-77.089005 38.913574, -77.029953 38.913574, -77.029953 38.886321,
            -77.089005 38.886321, -77.089005 38.913574))
K: <http://www.opengis.net/def/crs/OGC/1.3/CRS84> 
   Polygon((-77.089005 38.913574, -77.029953 38.913574, -77.029953 38.886321,
            -77.089005 38.886321, -77.089005 38.913574))    
\end{verbatim}

Then, we test whether these two geometries, i.e.~their corresponding WKT literals, are geometrically equal. This ensures that a correct answer to this test means that the underlying system assumes \href{http://www.opengis.net/def/crs/OGC/1.3/CRS84}{CRS84} as the default spatial reference system for WKT literals which do not specify one explicitly.

\reqstart
%\begin{itemize}
    \item Coordinate tuples within \href{http://www.opengis.net/ont/geosparql#wktLiteral}{\texttt{geo:wktLiterals}} shall be interpreted using the axis order defined in the spatial reference system used.
%\end{itemize}    
\reqend    
    
In order to test requirement 12, we define two new geometries in the dataset: L and M, which represent the same point. Geometry L has a WKT literal which specifies the point using the \href{http://www.opengis.net/def/crs/OGC/1.3/CRS84}{CRS84} coordinate system, while geometry M uses the \href{http://www.opengis.net/def/crs/EPSG/0/4326}{EPSG:4326} coordinate system \cite{nicolai2008new}. Compared to one another, these coordinate systems use an inverted axis order:

\begin{verbatim}
L: <http://www.opengis.net/def/crs/OGC/1.3/CRS84> Point(-88.38  31.95)
M: <http://www.opengis.net/def/crs/EPSG/0/4326>   Point( 31.95 -88.38)
\end{verbatim}

In order to test whether the system interprets the axis order correctly, i.e. according to the spatial reference system, we test if the two geometries are equal based on the system under test.

\reqstart
%\begin{itemize}
    \item An empty RDFS Literal of type \href{http://www.opengis.net/ont/geosparql#wktLiteral}{\texttt{geo:wktLiteral}} shall be interpreted as an empty geometry.
%\end{itemize}
\label{req:13}
\reqend

We define two new geometries, H and I, for the purpose of testing requirement 13. Geometry H represents a \texttt{LineString} geometry which has a WKT literal, which is an empty string. Geometry I represents an explicitly defined empty \texttt{LineString} geometry:

\begin{verbatim}
H: 
I: LineString EMPTY
\end{verbatim}

Additionally, as most of the other geometries, these two geometries have a \texttt{Point} representation, as well. In the case of geometry H, it is again represented by an empty value of the WKT literal, while geometry I has an explicitly defined empty \texttt{Point} geometry in its WKT literal:

\begin{verbatim}
H: 
I: Point EMPTY
\end{verbatim}

The test then consists of two parts, where both check if the WKT literals of \texttt{H} and \texttt{I} are equal. The two parts refer to the separate testing of the equality of the \texttt{LineString} geometries and the \texttt{Point} geometries. Both parts should be correct in order for requirement 13 to be fulfilled and thus fully scored by the benchmark.

\reqstart
%\begin{itemize}
    \item Implementations shall allow the RDF property \href{http://www.opengis.net/ont/geosparql#asWKT}{\texttt{geo:asWKT}} to be used in SPARQL graph patterns.
%\end{itemize}
\reqend

We test requirement 14 by simply selecting the \href{http://www.opengis.net/ont/geosparql#asWKT}{\texttt{geo:asWKT}} value of geometry A and checking it against the expected literal value.

\reqstart
%\begin{itemize}
    \item All \href{http://www.opengis.net/ont/geosparql#gmlLiteral}{\texttt{geo:gmlLiterals}} shall consist of a valid element from the GML schema that implements a subtype of GM\_Object as defined in \cite{portele2007ogc}.
%\end{itemize}    
\reqend    
    
For the purpose of testing requirement 15, we select all the values of the \href{http://www.opengis.net/ont/geosparql#asGML}{\texttt{geo:asGML}} property, regardless of the RDF subject, and check whether all of them contain a valid \texttt{GM\_Object} subtype in the value and whether its datatype is \href{http://www.opengis.net/ont/geosparql#gmlLiteral}{\texttt{geo:gmlLiteral}}. The ordered list of results is then checked against the expected answers, which include all valid GML literals from the dataset.

\reqstart
%\begin{itemize}
    \item An empty \href{http://www.opengis.net/ont/geosparql#gmlLiteral}{\texttt{geo:gmlLiteral}} shall be interpreted as an empty geometry.
%\end{itemize}
\reqend

Similarly to requirement 13, we test compliance to requirement 16 by providing an empty string as a GML literal value in one geometry - geometry H, and an explicitly defined empty \texttt{LineString} in a GML literal - geometry I:

\begin{verbatim}
H: 
I: <LineString><posList></posList></LineString>
\end{verbatim}

Just like with requirement 13, here we use a \texttt{Point} representations, as well. In the case of geometry H, it is again represented by an empty value of the GML literal, while geometry I has an explicitly defined empty \texttt{Point} geometry in its GML literal:

\begin{verbatim}
H: 
I: <Point><pos></pos></Point>
\end{verbatim}

The test for requirement 16 consists of two parts, as well, where both check if the GML literals of H and I are equal. The two parts refer to the separate testing of the equality of the \texttt{LineString} geometries and the \texttt{Point} geometries. Both parts should be correct in order for requirement 16 to be fulfilled.

\reqstart
%\begin{itemize}
    \item Implementations shall document supported GML profiles.
%\end{itemize}
\reqend

Requirement 17 is the only non-technical requirement of the GeoSPARQL standard, and therefore cannot be automatically checked and tested. This is the only requirement omitted by the benchmark tests. To keep it simple, we assume that all GeoSPARQL implementations fulfill this requirement and provide proper documentation for supported GML profiles, which we believe to be a reasonable assumption.

\reqstart
%\begin{itemize}
    \item Implementations shall allow the RDF property \href{http://www.opengis.net/ont/geosparql#asGML}{\texttt{geo:asGML}} to be used in SPARQL graph patterns.
%\end{itemize}
\reqend

Similarly to requirement 14, we test requirement 18 by simply selecting the \href{http://www.opengis.net/ont/geosparql#asGML}{\texttt{geo:asGML}} value of geometry A and checking it against the expected literal value.

\reqstart
%\begin{itemize}
    \item Implementations shall support \href{http://www.opengis.net/def/function/geosparql/distance}{\texttt{geof:distance}}, \href{http://www.opengis.net/def/function/geosparql/buffer}{\texttt{geof:buffer}}, \href{http://www.opengis.net/def/function/geosparql/convexHull}{\texttt{geof\-:\-convex\-Hull}},~ \href{http://www.opengis.net/def/function/geosparql/intersection}{\texttt{geof:intersection}},~ \href{http://www.opengis.net/def/function/geosparql/union}{\texttt{geof:union}},~ \href{http://www.opengis.net/def/function/geosparql/difference}{\texttt{geof:difference}}, \href{http://www.opengis.net/def/function/geosparql/symDifference}{\texttt{geof:symDifference}}, \href{http://www.opengis.net/def/function/geosparql/envelope}{\texttt{geof:envelope}} and \href{http://www.opengis.net/def/function/geosparql/boundary}{\texttt{geof:boundary}} as SPARQL extension functions, consistent with the definitions of the corresponding functions (distance, buffer, convexHull, intersection, difference, symDifference, envelope and boundary respectively) in Simple Features \cite{iso119125}.
%\end{itemize}
\reqend

In order to test requirement 19, we use separate tests for the nine functions in question, i.e.~we check each function separately. To test the full compliance of each function, we run three sub-tests for them: (a) we test the function with geometry parameters which are expressed as WKT literals, (b) we test it with geometry parameters expressed as GML literals, and (c) we test it with a combination of WKT and GML literals. If the function uses a single parameter, we only use the (a) and (b) sub-tests. If it uses two parameters, we use the (a), (b) and (c) sub-tests, where (c) consists of two queries in which WKT is the first and GML is the second parameter of the function (denoted as WKT-GML), and vice-versa (denoted as GML-WKT). With this, the test for each function consists of either two sub-tests (WKT and GML), or of four sub-tests (WKT-WKT, GML-GML, WKT-GML and GML-WKT). This ensures that the compliance score for each function is thoroughly checked. The scoring details for these tests are presented in  \Cref{sec:benchmarkresults}.

With this, the entire test for requirement 19 consists of tests for the nine functions, each with two or four sub-tests, for a total of 28 SPARQL queries.

\reqstart
%\begin{itemize}
    \item Implementations shall support \href{http://www.opengis.net/def/function/geosparql/getSRID}{\texttt{geof:getSRID}} as a SPARQL extension function.
%\end{itemize}
\reqend

We test requirement 20 by using the \href{http://www.opengis.net/def/function/geosparql/getSRID}{\texttt{geof:getSRID}} function in two queries: one with the WKT literal of geometry A, and the other with the GML literal of geometry A. In both cases we check if the system correctly returns \href{http://www.opengis.net/def/crs/OGC/1.3/CRS84}{\texttt{http://www.opengis.net/def/crs/OGC/1.3/CRS84}} as an answer.

\reqstart
%\begin{itemize}
    \item Implementations shall support \href{http://www.opengis.net/def/function/geosparql/relate}{\texttt{geof:relate}} as a SPARQL extension function, consistent with the relate operator defined in Simple Features \cite{iso119125}.
%\end{itemize}
\reqend

For testing requirement 21, we use a relate operator which denotes the \texttt{contains} relation (expressed as \texttt{T*****FF*} in DE-9IM \cite{Strobl2017}), and test it on geometries A and B, where A contains B in the dataset. Given that the \href{http://www.opengis.net/def/function/geosparql/relate}{\texttt{geof:relate}} function uses two parameters, there are four queries for this test: WKT-WKT, GML-GML, WKT-GML and GML-WKT.

\reqstart
%\begin{itemize}
    \item Implementations shall support \href{http://www.opengis.net/def/function/geosparql/sfEquals}{\texttt{geof:sfEquals}}, \href{http://www.opengis.net/def/function/geosparql/sfDisjoint}{\texttt{geof:sfDisjoint}}, \href{http://www.opengis.net/def/function/geosparql/sfIntersects}{\texttt{geof:sfIntersects}}, \href{http://www.opengis.net/def/function/geosparql/sfTouches}{\texttt{geof:sfTouches}}, \href{http://www.opengis.net/def/function/geosparql/sfCrosses}{\texttt{geof:sfCrosses}}, \href{http://www.opengis.net/def/function/geosparql/sfWithin}{\texttt{geof:sfWithin}}, \href{http://www.opengis.net/def/function/geosparql/sfContains}{\texttt{geof:sfContains}}, \href{http://www.opengis.net/def/function/geosparql/sfOverlaps}{\texttt{geof:sfOverlaps}} as SPARQL extension functions, consistent with their corresponding DE-9IM intersection patterns \cite{Strobl2017}, as defined by Simple Features \cite{iso119125}.
    \item Implementations shall support \href{http://www.opengis.net/def/function/geosparql/ehEquals}{\texttt{geof:ehEquals}}, \href{http://www.opengis.net/def/function/geosparql/ehDisjoint}{\texttt{geof:ehDisjoint}}, \href{http://www.opengis.net/def/function/geosparql/ehMeet}{\texttt{geof:ehMeet}},~ \href{http://www.opengis.net/def/function/geosparql/ehOverlap}{\texttt{geof:ehOverlap}},~ \href{http://www.opengis.net/def/function/geosparql/ehCovers}{\texttt{geof:ehCovers}},~ \href{http://www.opengis.net/def/function/geosparql/ehCoveredBy}{\texttt{geof:ehCoveredBy}}, \href{http://www.opengis.net/def/function/geosparql/ehInside}{\texttt{geof:ehInside}}, \href{http://www.opengis.net/def/function/geosparql/ehContains}{\texttt{geof:ehContains}} as SPARQL extension functions, consistent with their corresponding DE-9IM intersection patterns, as defined by Simple Features \cite{iso119125}.
    \item Implementations shall support \href{http://www.opengis.net/def/function/geosparql/rcc8eq}{\texttt{geof:rcc8eq}}, \href{http://www.opengis.net/def/function/geosparql/rcc8dc}{\texttt{geof:rcc8dc}}, \href{http://www.opengis.net/def/function/geosparql/rcc8ec}{\texttt{geof\-:\-rcc8ec}}, \href{http://www.opengis.net/def/function/geosparql/rcc8po}{\texttt{geof:rcc8po}}, \href{http://www.opengis.net/def/function/geosparql/rcc8tppi}{\texttt{geof:rcc8tppi}}, \href{http://www.opengis.net/def/function/geosparql/rcc8tpp}{\texttt{geof:rcc8tpp}}, \href{http://www.opengis.net/def/function/geosparql/rcc8ntpp}{\texttt{geof:rcc8ntpp}}, \href{http://www.opengis.net/def/function/geosparql/rcc8ntppi}{\texttt{geof:rcc8ntppi}} as SPARQL extension functions, consistent with their corresponding DE-9IM intersection patterns \cite{Strobl2017} , as defined by Simple Features \cite{iso119125}.
%\end{itemize}
\reqend

We test requirements 22, 23 and 24 by applying a separate set of tests for each of the twenty-four functions specified. Each function is tested by employing four queries: one with two WKT literals (WKT-WKT), one with two GML literals (GML-GML), and two with a combination of WKT and GML literals (WKT-GML and GML-WKT). Each of the queries tests if the relation implemented by the tested function is correct for the used geometries from the dataset, and each of them returns a \href{http://www.w3.org/2001/XMLSchema#boolean}{\texttt{xsd:boolean}} answer. The geometries used for the tests of each function are carefully selected in order to provide an unambiguous assessment of whether the function is supported and correctly implemented in the system under test.

\reqstart
%\begin{itemize}
    \item Basic graph pattern matching shall use the semantics defined by the RDFS Entailment Regime \cite{glimm2013sparql}.
%\end{itemize}
\reqend

For the purpose of testing requirements 25, 26 and 27, we use queries which require the system to select both materialized RDF triples, as well as inferred RDF triples, based on the specifics of each requirement. 

Therefore, we test requirement 25 using three separate queries: the first one selects all instances of the \href{http://www.opengis.net/ont/geosparql#Feature}{\texttt{geo:Feature}} class, where we expect the system to select instances of the subclasses of the class, as well, e.g.~\href{http://example.org/ApplicationSchema#PlaceOfInterest}{\texttt{my:PlaceOfInterest}}; the second and the third one select all instances with the \href{http://www.opengis.net/ont/geosparql#hasGeometry}{\texttt{geo:hasGeometry}} and \href{http://www.opengis.net/ont/geosparql#hasDefaultGeometry}{\texttt{geo:hasDefaultGeometry}} properties, but expect the results to contain entities which use subproperties of these properties, as well, e.g.~\href{http://example.org/ApplicationSchema#PlaceOfInterest}{\texttt{my:hasExactGeometry}}. 

\reqstart
%\begin{itemize}
    \item Implementations shall support graph patterns involving terms from an RDFS/OWL \cite{mcguinness2004owl} class hierarchy of geometry types consistent with the one in the specified version of Simple Features \cite{iso119125}.
%\end{itemize}
\reqend

For requirement 26, we use two separate queries: they select all instances of \href{http://www.opengis.net/ont/sf#Surface}{\texttt{sf:Surface}} and \href{http://www.opengis.net/ont/sf#Curve}{\texttt{sf:Curve}}, respectively, but expect the results to contain all instances of their subclasses, as well, such as \href{http://www.opengis.net/ont/sf#LineString}{\texttt{sf:LineString}} and \href{http://www.opengis.net/ont/sf#Polygon}{\texttt{sf:Polygon}}.

\reqstart
%\begin{itemize}
    \item Implementations shall support graph patterns involving terms from an RDFS/OWL class hierarchy of geometry types consistent with the GML schema that implements \texttt{GM\_Object} using the specified version of GML \cite{portele2007ogc}.
%\end{itemize}
\reqend

To test requirement 27, we use a single query which selects all instances of \href{http://www.opengis.net/ont/gml#Surface}{\texttt{gml:Surface}}, but the expected results include all instances of its subclass, \href{http://www.opengis.net/ont/gml#LineString}{\texttt{gml:LineString}}. 

\reqstart
%\begin{itemize}
    \item Basic graph pattern matching shall use the semantics defined by the RIF Core Entailment Regime [W3C SPARQL Entailment] for the RIF rules \cite{boley2010rif} \href{http://www.opengis.net/def/rule/geosparql/sfEquals}{\texttt{geor:sfEquals}}, \href{http://www.opengis.net/def/rule/geosparql/sfDisjoint}{\texttt{geor:sfDisjoint}}, \href{http://www.opengis.net/def/rule/geosparql/sfIntersects}{\texttt{geor:sfIntersects}}, \href{http://www.opengis.net/def/rule/geosparql/sfTouches}{\texttt{geor:sfTouches}}, \href{http://www.opengis.net/def/rule/geosparql/sfCrosses}{\texttt{geor:sfCrosses}}, \href{http://www.opengis.net/def/rule/geosparql/sfWithin}{\texttt{geor:sfWithin}}, \href{http://www.opengis.net/def/rule/geosparql/sfContains}{\texttt{geor:sfContains}}, \href{http://www.opengis.net/def/rule/geosparql/sfOverlaps}{\texttt{geor:sfOverlaps}}.
    \item Basic graph pattern matching shall use the semantics defined by the RIF Core Entailment Regime [W3C SPARQL Entailment] for the RIF rules \cite{boley2010rif} \href{http://www.opengis.net/def/rule/geosparql/ehEquals}{\texttt{geor:ehEquals}}, \href{http://www.opengis.net/def/rule/geosparql/ehDisjoint}{\texttt{geor:ehDisjoint}}, \href{http://www.opengis.net/def/rule/geosparql/ehMeet}{\texttt{geor:ehMeet}}, \href{http://www.opengis.net/def/rule/geosparql/ehOverlap}{\texttt{geor\-:\-eh\-Over\-lap}}, \href{http://www.opengis.net/def/rule/geosparql/ehCovers}{\texttt{geor:ehCovers}}, \href{http://www.opengis.net/def/rule/geosparql/ehCoveredBy}{\texttt{geor:ehCoveredBy}}, \href{http://www.opengis.net/def/rule/geosparql/ehInside}{\texttt{geor:ehInside}}, \href{http://www.opengis.net/def/rule/geosparql/ehContains}{\texttt{geor\-:\-eh\-Contains}}.
    \item Basic graph pattern matching shall use the semantics defined by the RIF Core Entailment Regime [W3C SPARQL Entailment] for the RIF rules \cite{boley2010rif} \href{http://www.opengis.net/def/rule/geosparql/rcc8eq}{\texttt{geor:rcc8eq}}, \href{http://www.opengis.net/def/rule/geosparql/rcc8dc}{\texttt{geor:rcc8dc}}, \href{http://www.opengis.net/def/rule/geosparql/rcc8ec}{\texttt{geor:rcc8ec}}, \href{http://www.opengis.net/def/rule/geosparql/rcc8po}{\texttt{geor:rcc8po}}, \href{http://www.opengis.net/def/rule/geosparql/rcc8tppi}{\texttt{geor:rcc8tppi}}, \href{http://www.opengis.net/def/rule/geosparql/rcc8tpp}{\texttt{geor:rcc8tpp}}, \href{http://www.opengis.net/def/rule/geosparql/rcc8ntpp}{\texttt{geor:rcc8ntpp}}, \href{http://www.opengis.net/def/rule/geosparql/rcc8ntppi}{\texttt{geor:rcc8ntppi}}.
%\end{itemize}
\reqend

We test the requirements 28, 29 and 30 with eight different queries each, in order to test the sub-requirements for each individual rule specified. The queries used here are similar to the queries for requirements 4, 5 and 6, with the difference that the tests for requirements 28, 29 and 30 require both materialized RDF triples and inferred RDF triples to be selected for the query response. To ensure that the system selects all such entities, and therefore supports the semantics defined in the RIF core entailment regime for the RIF rules, the tests require an ordered list of entities fulfilling the query request.

\subsection{Benchmark Results}
\label{sec:benchmarkresults}

The benchmark can test if the benchmarked system provides a correct or an incorrect answer on each of the 206 benchmark queries. In order to transform these individual results into an overall result, we calculate two benchmark results from a given experiment:

\begin{itemize}
    \item \textbf{Correct answers}: The number of correct answers out of all GeoSPARQL queries, i.e.~tests.
    \item \textbf{GeoSPARQL compliance percentage}: The percentage of compliance with the requirements of the GeoSPARQL standard.
\end{itemize}

The former is straightforward -- it's the number of correct answers the system provided, out of the 206 test queries. The latter is calculated from the perspective of the 30 requirements and measures the overall compliance of the benchmarked system with the GeoSPARQL standard. It measures the amount of supported requirements of the system, out of the 30 specified requirements, where the weight of each requirement is uniformly distributed, i.e.~each requirement contributes 3.33\% to the total result.

If a requirement contains multiple sub-test queries, its 3.33\% are uniformly distributed among them. Therefore, for instance, each of the eight sub-require\-ments of requirement 4 contributes with 12.5\% to the parent test score, i.e.~with 0.4167\% (3.33\% x 12.5\%) to the total benchmark compliance percentage score. This means that a single requirement from the GeoSPARQL standard can be fully supported, partially supported or not supported at all.

The only exception of this rule of uniform distribution of the weights between tests on the same level, are the sub-test queries which test GeoSPARQL functions with different serializations of literals as parameters, i.e.~requirements 19 - 24. When we test a function for compliance to the standard while using (a) WKT-only literals, (b) GML-only literals and (c) a combination of WKT and GML literals, the score is uniformly distributed between these three logical groups, each contributing with 33.33\% to the parent test score. However, (c) is practically tested using two queries: one where WKT is the first and GML is the second parameter of the function (denoted as WKT-GML), and vice-versa (denoted as GML-WKT). These two queries technically contribute with 16.67\% to the parent test score each, so that the total contribution from the logical group (c) remains 33.33\%. With this, the technical weight of the queries themselves is 33.33\% for the WKT-only query, 33.33\% for the GML-only query, 16.67\% for the WKT-GML query, and 16.67\% for the GML-WKT query. Technically, on a query level, this is an exception of the uniform distribution rule we practice, but logically, on a group level, it still holds.

Given that requirement 17 is non-technical, and therefore not tested as part of the benchmark, each system gets its 3.33\% score points automatically, when it provides at least one correct answer to the benchmark tests.

\subsection{Benchmark Considerations}
\label{sec:considerations}

When creating the benchmark we needed to take certain considerations and interpretations which were implicitly given in the GeoSPARQL standard. We elaborate on these in this subsection.

\subsubsection*{Geometry Literals}

Many results of query functions defined in the GeoSPARQL standard return a \href{http://www.opengis.net/ont/geosparql#geomLiteral}{\texttt{ogc:geomLiteral}} as a result following the GeoSPARQL standard definition. This means that, according to the standard, a function such as:

%\href{http://www.opengis.net/def/function/geosparql/boundary}{\texttt{geof:boundary(ogc:geomLiteral):ogc:geomLiteral}}
\begin{verbatim}
   geof:boundary(ogc:geomLiteral):ogc:geomLiteral
\end{verbatim}
\noindent
may take either a WKT, a GML 2.0, or a GML 3.2 literal as an argument, and may return either a WKT, a GML 2.0, or a GML 3.2 literal as a result. The dataset we use for our benchmark includes WKT and GML 3.2 formatted literals. However, we provide query answers in WKT, GML 2.0 and GML 3.2 to support all possible outcomes from a system tested by the benchmark.

The decision to include only GML 3.2 and not GML 2.0 literals in our dataset was taken because GML 2.0 has been de-facto superseded by GML 3.2. GML 2.0 is not even supported as an export option in current GIS software, such as QGIS for instance. In addition, in all systems we benchmarked, the only GML variant that was supported was GML 3.2.

\subsubsection*{Variations Between Literal Serializations}

Within the same literal type, different semantically equivalent representations of geometries are possible. WKT serializations may include a CRS URI, but they may also omit it (if it's missing, WGS84 CRS is assumed), and they may differ in the amount and positioning of whitespaces. GML literals may differ in the order of attributes and definition of namespaces. To be flexible about these variations, we apply a normalization process before comparing the results from the tested system with the expected answer. WKT literals are trimmed and their whitespaces are removed, and GML literals are converted to canonicalized XML with normalized namespace definitions.

\subsubsection*{Alternative Answers}

The GeoSPARQL standard defines the results of GeoSPARQL functions as \href{http://www.opengis.net/ont/geosparql#geomLiteral}{\texttt{ogc:geomLiteral}} values, but does not define which geometry types these literals should serialize. Therefore, functions may not only return results in different literal types, but also in different geometry representations even within the same literal serialization. One example is the \href{http://www.opengis.net/def/function/geosparql/boundary}{\texttt{geof:boundary}} function which could return a \href{http://www.opengis.net/ont/sf#LinearRing}{\texttt{sf:LinearRing}} or a \href{http://www.opengis.net/ont/sf#Polygon}{\texttt{sf:Polygon}} geometry as a result. Even supposedly simple return values such as an \href{http://www.w3.org/2001/XMLSchema#boolean}{\texttt{xsd:boolean}} may be represented as either the \href{http://www.w3.org/2001/XMLSchema#boolean}{\texttt{xsd:boolean}} literals with value \emph{true} and \emph{false} or \emph{1} and \emph{0}. 

In order to deal with these scenarios, we define alternative query answers for each of the aforementioned possibilities. This means that each test consists of a single query which is issued to the system under test, and a set of several alternative correct answers, which are logically equivalent, but may be technically represented in different serializations.

\subsection{Implementation}
\label{sec:implementation}

\begin{figure}[!ht]
    \centering
    \includegraphics[width=0.9\textwidth]{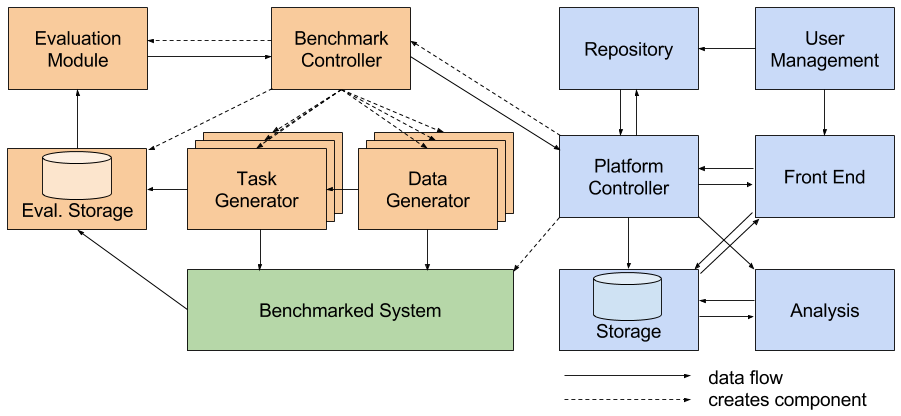}
    \caption{The HOBBIT benchmarking platform.}
    \label{fig:hobbitplatformdiagram}
\end{figure}

We have implemented the benchmark as a benchmark for the HOBBIT platform\footnote{Public instance of the HOBBIT Platform: \href{http://master.project-hobbit.eu}{http://master.project-hobbit.eu}}, intended for holistic benchmarking of big Linked Data \cite{roder2019hobbit}. The HOBBIT platform allows for users to define and execute benchmarks, on one hand, and provide and add triplestore systems, on the other. A user can run an experiment on the platform by selecting the desired benchmark and the target triplestore system to be tested. The platform then loads the benchmark as a set of Docker containers (benchmark controller, data generator, task generator and evaluation module), loads the system as a Docker container (benchmarked system), and then runs the benchmark according to its logic, programmed in the controller (\Cref{fig:hobbitplatformdiagram}). The results of each experiment are stored in the platform and are made publicly available on the Web.

In our case, the GeoSPARQL compliance benchmark first loads the dataset into the benchmarked system, then reads all the test queries and sends them to the benchmarked system for execution. The evaluation module reads the single expected answer or the set of expected alternative answers for each query, and compares whether the benchmarked system returns a correct or an incorrect answer, saving the result into the evaluation store. After all tests are done, the evaluation module calculates two summarized results: (1) the number of correct answers, out of all possible tests, and (2) the percentage of compliance to the requirements of the GeoSPARQL standard, as described in \Cref{sec:benchmarkresults}.

We decided to use the HOBBIT platform for our benchmark due to its plug-in nature, in which additional systems can be added by interested users, which will then be able to run an experiment with the benchmark over their own system. A user can also run our GeoSPARQL compliance benchmark over any triplestore system which is already available on the platform. Additionally, the public nature of the platform allows for greater transparency and reproducibility of the results of each benchmark, including our GeoSPARQL compliance benchmark.

\section{Experimental Setup}
\label{sec:experimentalsetup}

In order to showcase the usability and usefulness of the GeoSPARQL compliance benchmark, we set out to run a number of experiments over some of the most commonly used triplestores. The set of chosen triplestores is shown in \Cref{tab:triplestoretests}.

\setlength{\tabcolsep}{0.5em}
\begin{table}[!h]
    \centering
    \begin{tabular}{c|c|c}
    \toprule
        \textbf{Triplestore} & \textbf{Version} & \textbf{Reference}\\
        \midrule
        Apache Marmotta & 3.4.0 & \cite{apachemarmotta} \\
        Blazegraph & 3.1.5 & \cite{blazegraph} \\
        Eclipse RDF4J & 3.4.0 & \cite{rdf4j} \\
        Jena Fuseki & 3.14.0 & \cite{jenafuseki} \\
        GeoSPARQL Fuseki & 3.17.0 & \cite{albiston2018geosparql, geosparqlfuseki}\\
        Ontotext GraphDB & 9.3.3 & \cite{graphdb} \\ 
        OpenLink Virtuoso & 7.3 & \cite{erling2012virtuoso, virtuoso} \\
        % Parliament & 2.7.10 \\ 
        Stardog & 7.4.0 & \cite{stardog} \\
        \bottomrule
    \end{tabular}
    \vspace{10pt}
    \caption{Triplestores which have been tested using the GeoSPARQL compliance benchmark.}
    \label{tab:triplestoretests}
\end{table}

For each experiment, a system adapter has been created and published on a public HOBBIT platform instance, as well as in the HOBBIT GitLab repository\footnote{HOBBIT Platform Triplestores: \href{https://git.project-hobbit.eu/triplestores}{https://git.project-hobbit.eu/triplestores}}. This allows for the reproduction of the experiments and the results. 

Each triplestore version from \Cref{tab:triplestoretests} was the most recent available stable version of the implementation at the time of testing. For each of the triplestores which have been tested, a system adapter implementation has been created which handles the initial configuration of the triplestore, e.g.~setting up a repository which contains the data to be tested, enabling geospatial query support, etc. If possible, this adapter implementation was added to the triplestore implementation in a joint Docker image or two Docker images -- the adapter implementation and the triplestore implementation -- were created for testing. It needs to be stated that not all of the aforementioned triplestores claim to support GeoSPARQL. In fact, Blazegraph and Jena Fuseki do not support GeoSPARQL. We included them in our experiments in order to show which GeoSPARQL requirements are already supported by a non-GeoSPARQL implementation of an RDF triplestore.

\section{Results and Discussion}
\label{sec:results}

\subsection{Overall Results}

\begin{figure}[!ht]
    \centering
    \includegraphics[width=\textwidth]{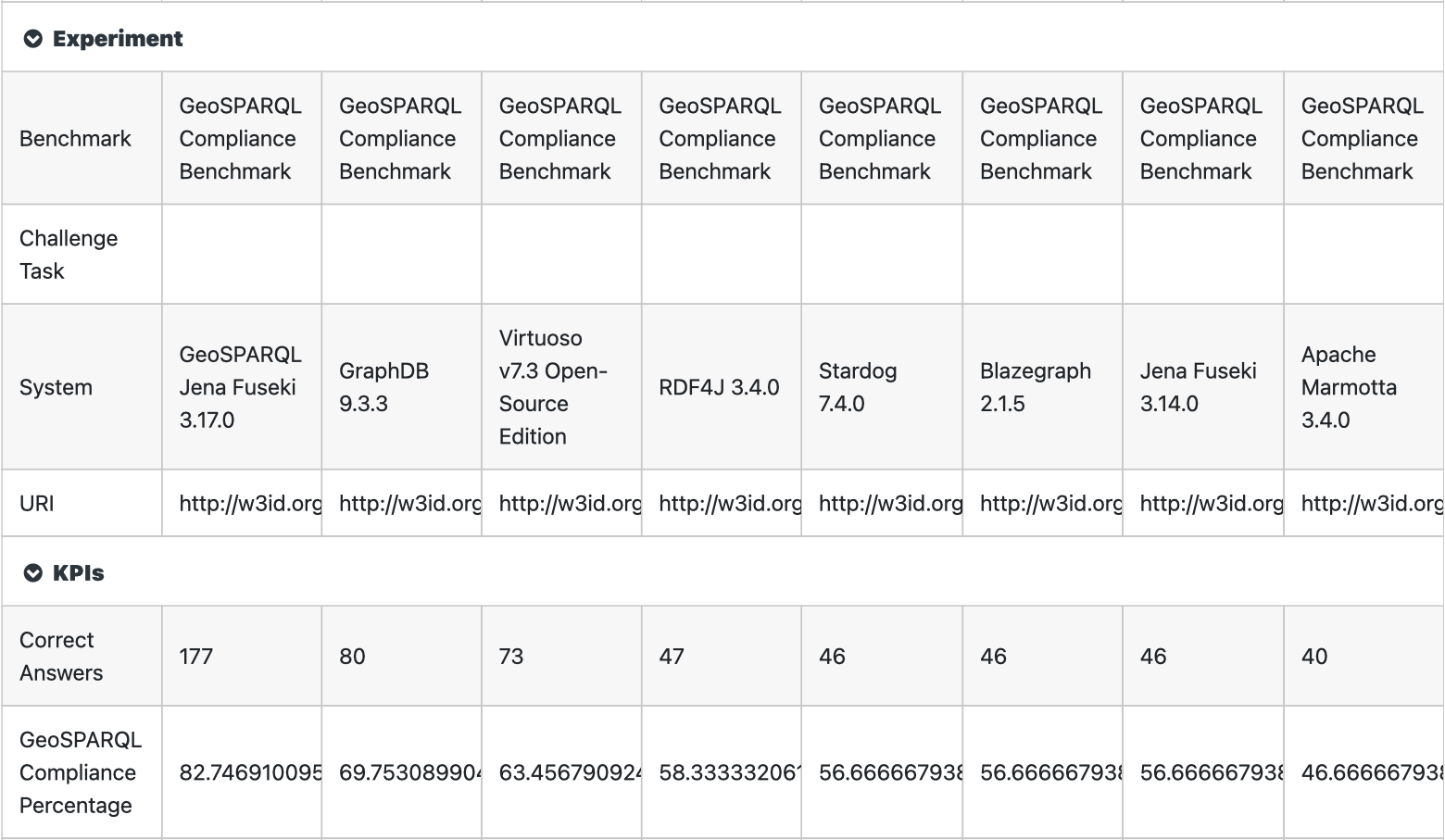}
    \caption{Results from the GeoSPARQL compliance benchmark, from the public instance of the HOBBIT platform.}
    \label{fig:compliancebnchmarkresults}
\end{figure}

The results of the experiments with our benchmark and the systems listed in \Cref{tab:triplestoretests} are shown in \Cref{tab:results} and on \Cref{fig:compliancebnchmarkresults}, and are available online on the HOBBIT platform\footnote{Results on the HOBBIT platform: \href{https://master.project-hobbit.eu/experiments/1612476122572,1612477003063,1612476116049,1612477500164,1612661614510,1612637531673,1612828110551,1612477849872}{https://master.project-hobbit.eu/experiments/1612476122572,1612477003063,1612476116049,\\1612477500164,1612661614510,1612637531673,1612828110551,1612477849872}}. They show that none of these widely used RDF storage solutions fully comply to the GeoSPARQL standard. Aside from that, we can point out that one of them stands out with a significantly better GeoSPARQL compliance score than the others, and more generally, the top three stand out from the rest. The triplestores in positions 4 - 7 share an almost identical result.

% \setlength{\tabcolsep}{0.5em}
% \begin{table}[htb]
%     \centering
%     \begin{tabular}{c|c|c|c|c}
%         \toprule
%         \textbf{Triplestore} & \textbf{Correct Answers} & \textbf{Benchmark Result} & \textbf{Geo} & \textbf{Geo+RDFS} \\
%         \midrule
%         GeoSPARQL Fuseki 3.17 & 177 & 82.75\% & 59.2\% &  66.4\% \\
%         Ontotext GraphDB 9.3.3 & 80 & 69.75\% & 29.75\% & 42.5\% \\ 
%         OpenLink Virtuoso 7.3 & 73 & 63.46\% &  15\% & 30.9\% \\
%         Eclipse RDF4J 3.4.0 & 47 & 58.33\% & 3.7\% & 21.4\% \\
%         Blazegraph 2.1.5 & 46 & 56.67\% & 0\% & 18.4\% \\
%         Stardog 7.4.0 & 46 & 56.67\% & 0\% &  18.4\% \\
%         Jena Fuseki 3.14 & 40 & 46.67\% & 0\% &  18.4\% \\
%         Apache Marmotta 3.4.0 & 40 & 46.67\% & 0\% &  18.4\% \\
%         % Parliament 2.7.10 & 0 & 0\% \\
%         \bottomrule
%     \end{tabular}
%     \caption{Results from the GeoSPARQL compliance benchmark. The first colum describes the correct answers by the respective implementation, the second column, the GeoSPARQL compliance score. The third column \emph{Geo} describes a normalized score of all GeoSPARQL extensions excluding CORE,TOP and RDFSE which should be supported by any SPARQL and RDFS compliant triplestore. The last column \emph{Geo+RDFS} describes a normalized score of all extensions excluding CORE and TOP.}
%     \label{tab:results}
% \end{table}

\setlength{\tabcolsep}{0.5em}
\begin{table}[htb]
    \centering
    \begin{tabular}{c|c|c}
        \toprule
        \textbf{Triplestore} & \textbf{Correct Answers} & \textbf{GeoSPARQL Compliance} \\
        \midrule
        GeoSPARQL Fuseki 3.17 & 177 & 82.75\% \\
        Ontotext GraphDB 9.3.3 & 80 & 69.75\% \\ 
        OpenLink Virtuoso 7.3 & 73 & 63.46\% \\
        Eclipse RDF4J 3.4.0 & 47 & 58.33\% \\
        Stardog 7.4.0 & 46 & 56.67\% \\
        Blazegraph 2.1.5 & 46 & 56.67\% \\
        Jena Fuseki 3.14 & 46 & 56.67\% \\
        Apache Marmotta 3.4.0 & 40 & 46.67\% \\
        % Parliament 2.7.10 & 0 & 0\% \\
        \bottomrule
    \end{tabular}
    \vspace{10pt}
    \caption{Results from the GeoSPARQL compliance benchmark.}
    \label{tab:results}
\end{table}

In order to see the reasons for these variations more closely, we made a breakdown of the compliance results into the six extensions defined in the GeoSPARQL standard. These results are shown in \Cref{tab:resultsextesions}. As we can see from this table, the triplestores in positions 4 - 7 share the same result due to demonstrating full compliance with the CORE, TOP and RDFSE extensions of the GeoSPARQL benchmark, but not with the other extensions. The reason why almost all benchmarked triplestores comply with CORE, TOP and RDFSE is simple: these requirements are designed in such a way that they are satisfied ``out-of-the-box'' by most RDF- and SPARQL-compliant storage solutions. They refer to the use of specific classes (CORE) and properties (TOP) in SPARQL query patterns, as well as RDFS reasoning (RDFSE), which are features supported in most triplestores nowadays. Since RDFS reasoning was not activated in the Marmotta version we benchmarked, it has no compliance for RDFSE so its score comes only from its compliance with CORE and TOP, thus is lower than the scores of the other systems.

The bottom three systems are explicitly not GeoSPARQL-compliant, but we included them in our experiments as baseline tests. As we can see, they all demonstrated compatibility with either two or with three extensions of the GeoSPARQL standard (\Cref{tab:resultsextesions}), and scored 56.67\% or 46.67\% of the GeoSPARQL compliance score (\Cref{tab:results}). This, however, does not mean that the benchmark score should start at 56.67\% or 46.67\%, since a benchmarked RDF storage system may fail these tests, too.

\begin{table}[!h]
    \centering
    \begin{tabular}{c|c|c|c|c|c|c}
        \toprule
        \textbf{Triplestore} & \textbf{CORE} & \textbf{TOP} & \textbf{GEOEXT} & \textbf{GTOP} & \textbf{RDFSE} & \textbf{QRW}\\
        \midrule
        GeoSPARQL Fuseki & \emph{Full} & \emph{Full} & \emph{Full/E} & \emph{Full} & \emph{Full} & \emph{Full/E} \\
        Ontotext GraphDB & \emph{Full} & \emph{Full} & \emph{Partial [WKT]} & \emph{Partial [WKT]} & \emph{Full} & \emph{None} \\ 
        OpenLink Virtuoso & \emph{Full} & \emph{Full} &\emph{Partial [WKT]} & \emph{Partial [WKT]} & \emph{Full} & \emph{None} \\
        Eclipse RDF4J & \emph{Full} & \emph{Full} & \emph{Partial [WKT CRS84]} & \emph{Partial [WKT CRS84]} & \emph{Full} & \emph{None} \\
        Stardog & \emph{Full} & \emph{Full} & \emph{None} & \emph{None} & \emph{Full} & \emph{None} \\
        Blazegraph & \emph{Full} & \emph{Full} & \emph{None} & \emph{None} & \emph{Full} & \emph{None} \\
        Jena Fuseki & \emph{Full} & \emph{Full} & \emph{None} & \emph{None} & \emph{Full} & \emph{None} \\
        Apache Marmotta & \emph{Full} & \emph{Full} & \emph{None} & \emph{None} & \emph{None} & \emph{None} \\
        % Parliament & \emph{None} & \emph{None} & \emph{None} & \emph{None} & \emph{None} & \emph{None} \\
        \bottomrule
    \end{tabular}
    \vspace{10pt}
    \caption{Support of the different GeoSPARQL extension by the tested triplestores. \emph{Full} indicates full support, comprised of correct query answers only, \emph{Full/E} indicates that support is implemented but erroneous, \emph{Partial [GML/WKT]} indicates that support is partially implemented, \emph{None} indicates that support for this GeoSPARQL extension is not present.}
    \label{tab:resultsextesions}
\end{table}

\subsection{Discussion on the Results for each Triplestore}

First, we tested RDF triplestores which claim GeoSPARQL support. We wanted to check how extensive their compliance with the GeoSPARQL benchmark is, and this list included: GeoSPARQL Fuseki, GraphDB, Virtuoso, RDF4J and Stardog.

GeoSPARQL Fuseki is the triplestore with the highest GeoSPARQL compliance score in our experiments. It is the only system with full GML and WKT support and the only system with a full implementation of all GeoSPARQL extensions (\Cref{tab:resultsextesions}). However, GeoSPARQL Fuseki produced incorrect results in many functions covered by the query rewrite extension and in a few functions covered by the geometry extension. Also, just like all other triplestores we tested, GeoSPARQL Fuseki fails to handle empty WKT and empty GML literals. 

GraphDB provides a full implementation of all but the query rewrite extension. However, GraphDB can only handle WKT literals, but not GML literals. This leads to a substantially lower score in our benchmark, as many queries require either a GML literal as input, or a combination of a GML and a WKT literal in order to be executed. Most functions with WKT-only literals in the GEOEXT and GTOP extension tests produced correct results.

Virtuoso provides support for WKT literals, but not GML literals. Similarly to GraphDB, it provides full implementation for all GeoSPARQL extensions, except for the query rewrite extension. However, it has an additional issue: even though it returns logically correct results for the tests for the functions in requirement 19 (part of the GEOEXT extension), the literals are transformed from WKT literals to an internal literal type which is Virtuoso-specific. This renders a mismatch between the provided and expected answer, and lowers the benchmark score for Virtuoso.

The RDF4J triplestore implements all the GeoSPARQL functions of the GEOEXT extension and the GTOP extension for WKT literals. However, RDF4J fails almost all of the GeoSPARQL tests from these extensions because it does not support CRS URIs in WKT literals. While the GeoSPARQL standard acknowledges that the integration of CRS URIs in WKT Literals is optional, they are used in various use-cases, especially at geospatial authorities, and we expect them to be supported in every triplestore which claims GeoSPARQL support. Thus, WKT literals with explicit CRS URIs are included in most of the tests of the benchmark. In addition, RDF4J lacks support for GML literals and the query rewrite extension.

The Stardog triplestore provides an implementation covering WKT literals and implements five geospatial functions which are similar to the GeoSPARQL functions, but not fully compatible. More specifically, out of their five geospatial functions (\texttt{geof:within}, \texttt{geof:area}, \texttt{geof:nearby}, \texttt{geof:distance} and \texttt{geof:relate}), only the \texttt{\href{http://www.opengis.net/def/function/geosparql/distance}{geof:distance}} function follows the signature of the GeoSPARQL function with the same URI. However, our tests for this function include WKT literals with explicit CRS URIs, which Stardog doesn't support, so the test for this function fails. The tests for the other functions fail either because functions with those URIs don't exist in the GeoSPARQL standard, or because of a function signature mismatch. Thus, Stardog only scores in tests which cover the CORE, TOP and RDFSE extensions. 

%claims the implementation of ``five of the major operators defined by GeoSPARQL''\footnote{\url{https://www.stardog.com/blog/geospatial-a-primer/}}. Out of the five functions which are claimed to be supported (\href{http://www.opengis.net/def/function/geosparql/within}{geof:within}, \href{http://www.opengis.net/def/function/geosparql/area}{geof:area}, \href{http://www.opengis.net/def/function/geosparql/nearby}{geof:nearby}, \href{http://www.opengis.net/def/function/geosparql/distance}{geof:distance} and \href{http://www.opengis.net/def/function/geosparql/relate}{geof:relate}), only the \href{http://www.opengis.net/def/function/geosparql/distance}{geof:distance} function follows the GeoSPARQL signature of the function. The signature of \href{http://www.opengis.net/def/function/geosparql/relate}{geof:relate} does not match with the one defined in GeoSPARQL, \href{http://www.opengis.net/def/function/geosparql/within}{geof:within}'s signature matches, but is called \href{http://www.opengis.net/def/function/geosparql/sfWithin}{geof:sfWithin} in GeoSPARQL and the functions \href{http://www.opengis.net/def/function/geosparql/area}{geof:area} and \href{http://www.opengis.net/def/function/geosparql/nearby}{geof:nearby} are not defined in the GeoSPARQL standard. When testing the geof:distance function we realized that it cannot cope with CRS URIs in WKT literals. Thus, Stardog only scores in tests which cover the CORE, TOP and RDFSE extensions. 

Next, we tested triplestores which do not claim to support GeoSPARQL, but claim support for other geospatial extensions. We expected that they will provide full support for the GeoSPARQL CORE, TOP and RDFSE extensions which do not rely on the implementation of additional geospatial operators. They thereby constitute as baseline tests for our approach, and this list included: Blazegraph, Jena Fuseki, Apache Marmotta and Parliament.

Blazegraph supports some non-GeoSPARQL spatial functions in its GeoSpatial Search Extension\footnote{Blazegraph GeoSpatial: \url{https://github.com/blazegraph/database/wiki/GeoSpatial}}. This extension allows the definition of \texttt{Points} via WKT literals, but is otherwise not GeoSPARQL-compliant. Blazegraph therefore fails the GEOEXT, GTOP and QRW tests, as expected.

Jena Fuseki includes a customized spatial extension \emph{Jena Spatial}\footnote{Jena Spatial: \url{https://jena.apache.org/documentation/query/spatial-query.html}} which is planned to be replaced by the GeoSPARQL Fuseki implementation we tested. Jena Fuseki can cope with WKT literals and defines a custom set of functions, none of which match the function signatures defined in the GeoSPARQL standard. Hence, Jena Fuseki only gets awarded a full score in the CORE, GTOP and RDFSE extensions.

Apache Marmotta has a GeoSPARQL implementation which was created in a Google Summer of Code project\footnote{Marmotta GeoSPARQL: \url{http://marmotta.apache.org/kiwi/geosparql.html}}. At the time of testing the extension was not included in the last stable version of this triplestore, therefore the version we tested was not GeoSPARQL-compliant. Even though Marmotta supports RDFS reasoning, we were unsuccessful in our attempts to activate it on the instance we worked with, so even though we expected it to achieve the same score as the other triplestores which do not support GeoSPARQL, it only scored as compliant with CORE and TOP. %We expect the GeoSPARQL compliance score to increase with the release of Apache Marmotta 3.5.0 which is supposed to include the GeoSPARQL extension.

Finally, we want to acknowledge that we also tested the Parliament 2.7.10 triplestore. Parliament validates WKT and GML literals before they are added to the graph, and fails to load a dataset if a validation error occurs. In our test, Parliament failed to parse GML 3.2 literals and the empty WKT literals. As a result, the benchmark dataset could not be loaded and we could not conduct the experiment with the Parliament triplestore.

\section{Limitations of the Benchmark}
\label{sec:limitations}

% MILOS (10.02.2021): We remove this paragraph for now. We can work on improving Req. 10 and 15 in the future. % The GeoSPARQL benchmark tests the compliance of the requirements stated in the GeoSPARQL specification. However, the GeoSPARQL specification document refers to several further standards which this benchmark does not cover. For instance, the benchmark does not check the validity of all possible types of geometries which could be represented in WKT or GML literals and it does not check if a triplestore deals with incorrectly formatted literals of both kinds. We figured that testing these capabilities would be out of scope for this benchmark.

The GeoSPARQL compliance benchmark does not test every GeoSPARQL function with every available geometry type and their combinations. We do that with WKT and GML serializations, but not different geometry types. The reason for this is that the amount of possible combinations of geometries would be inconceivably too large and the benefit of testing them far too low. WKT defines 27 geometry types, GML defines at least as many which would need to be considered in both in their GML 2.0 and in their GML 3.2 variants, to be complete. Instead, our dataset consists of \texttt{Points}, \texttt{LineStrings} and \texttt{Polygons}, which are the most widely used geometry types. With this, we believe we strike a good balance between the benchmark being too extensive and being sufficiently precise in measuring a system's compliance with the GeoSPARQL standard.

Regarding the GeoSPARQL compliance percentage score: as we already stated, this score measures the number of supported requirements of the system, out of the 30 specified requirements, where the weight of each requirement is uniformly distributed, i.e.~each requirement contributes 3.33\% to the total result. The reason we decided to use uniform distribution instead of assigning requirement-specific weights, is because adding weights to different requirements would be somewhat arbitrary. Given that the authors of the GeoSPARQL standard have not discussed or put any variable significance between the different requirements, gives us a signal that, at least for the time being, we should treat them as equally important. While that practically isn't the case, and different stakeholders may have different significance implicitly assigned to them, we don't think there is a better universal way to address this.

% Another argument against our scoring might be the fact that even non-GeoSPARQL triplestores score high with the benchmark. However, given that the requirements from the CORE, TOP and RDFSE extensions work in most RDF storage systems, they are still official requirements defined in the GeoSPARQL standard, and a system failing to support some or all of these, would definitely not comply with the requirements from the other extensions of the standard. We should therefore include it in our score. We must also make note that we tested some of the most mature RDF triplestores, so we expected their results to be high. Additionally, one of the tested triplestores discussed in \Cref{sec:results} scored 0\% due to not being able to load an RDF dataset with empty WKT and GML literals.

\section{Conclusions}
\label{sec:conclusions}

This paper introduces a GeoSPARQL compliance benchmark which aims to measure the extent to which an RDF triplestore complies with the requirements specified in the GeoSPARQL standard. By doing a series of tests for each requirement, the benchmark is able to assess whether the benchmarked system fully or partially supports a given requirement, or not at all. The results from the 206 individual tests are transformed into a GeoSPARQL compliance percentage which aims to provide a metric of the amount of requirements covered by the benchmarked system.

In order to showcase the usefulness and usability of the benchmark, as part of the HOBBIT platform, we ran a series of experiments with eight of the most commonly used RDF triplestores. The overall results show that GeoSPARQL support varies greatly between the tested triplestores. While the CORE, TOP and RDFSE extensions are supported in almost every triplestore -- as they only depend on SPARQL and RDFS functionalities and are not GeoSPARQL-specific -- the GEOEXT and GTOP extensions show varying levels of implementation. Some triplestores, such as GraphDB or Virtuoso, chose to only implement support for WKT literals, RDF4J supports only WKT literals without CRS URIs and only GeoSPARQL-Jena provides a full GeoSPARQL-compliant implementations of all functions with both GML and WKT compatibility. GeoSPARQL-Jena is also the only implementation tested in our benchmark which implements the QRW extension of GeoSPARQL.

In conclusion, we can see that the GeoSPARQL standard, almost nine years after its initial release, is often only partially supported by major triplestore vendors. We hope that the contribution of our GeoSPARQL benchmark can help to motivate implementers to improve their RDF storage solutions, give customers a guideline as to which implementation is most suitable for their given use-case, and provide a starting point for a further standard-conform expansion of the geospatial Semantic Web.

\subsection{Future Work}
\label{sec:futurework}

Recently, the OGC GeoSPARQL Working Group has been reactivated \cite{homburg2020ogc,homburg2020ogc2} to define GeoSPARQL 2.0, a successor to the GeoSPARQL standard. It is a good practice of emerging OGC standards to first be defined, then reviewed, and at the same time also implemented as a proof-of-concept. During the course of this implementation, compliance testing becomes increasingly common as can be seen by the establishment of the OGC Team Engine\footnote{OGC Team Engine: \href{https://cite.opengeospatial.org/teamengine/}{https://cite.opengeospatial.org/teamengine/}}, a compliance test suite which enterprises may use to get official OGC compliance certifications for their software implementations. Given that currently no OGC-endorsed OGC GeoSPARQL compliance test exists, we would welcome a collaboration with the OGC and would like to extend our test suite to cover the changes which will be defined in GeoSPARQL 2.0.

\section*{Acknowledgement}
\label{sec:acknowledgement}

This work has been partially supported by Eurostars Project SAGE (GA no. E!10882).

%\bibliographystyle{unsrt}
%\bibliography{bibliography}

\end{document}